\documentclass{article}
\usepackage{ijcai21}

\usepackage{times}
\usepackage{soul}
\usepackage{url}
\usepackage[hidelinks]{hyperref}
\usepackage[utf8]{inputenc}
\usepackage[small]{caption}
\usepackage{graphicx}
\usepackage{amsmath}
\usepackage{amsthm}
\usepackage{booktabs}
\usepackage{subfig}
\usepackage{algorithm}
\usepackage{algorithmicx}
\usepackage[noend]{algpseudocode}
\usepackage{multicol}
\usepackage{tabularx} 
\usepackage{amssymb}
\usepackage{verbatim}
\usepackage{multirow}
\usepackage{graphbox}
\title{Stochastic Market Games}

\author{
Kyrill Schmid$^1$\and
Lenz Belzner$^2$\and
Robert Müller$^1$\and
Johannes Tochtermann $^1$ \And 
Claudia Linnhoff-Popien $^1$\\
\affiliations
$^1$LMU Munich \\
$^2$Technische Hochschule Ingolstadt	\\
\emails
\{kyrill.schmid, robert.mueller, linnhoff\}@ifi.lmu.de,
lenz.belzner@thi.de,
johannes.tochtermann@campus.lmu.de
}

\begin{document}

\maketitle

\begin{abstract}
Some of the most relevant future applications of multi-agent systems like autonomous driving or factories as a service display mixed-motive scenarios, where agents might have conflicting goals. In these settings agents are likely to learn undesirable outcomes in terms of cooperation under independent learning, such as overly greedy behavior. Motivated from real world societies, in this work we propose to utilize market forces to provide incentives for agents to become cooperative. As demonstrated in an iterated version of the Prisoner's Dilemma, the proposed market formulation can change the dynamics of the game to consistently learn cooperative policies. Further we evaluate our approach in spatially and temporally extended settings for varying numbers of agents. We empirically find that the presence of markets can improve both the overall result and agent individual returns via their trading activities.
\end{abstract}
\section{Introduction}\label{sec:introduction}
In addition to the breakthroughs of multi-agent learning in fully cooperative domains  \cite{foerster2018counterfactual,rashid2018qmix,sunehag2018value}, recently more attention has been paid to multi-agent learning in mixed scenarios, where multiple self-interested agents live in a common environment \cite{dafoe2020open,hostallero2020inducing}. In mixed scenarios agents are characterized by general sum returns, so agents might compete or even have conflicting goals. These settings give rise to severe problems such as the tragedy of the commons \cite{ostrom1990governing,hardin2009tragedy,perolat2017multi}, and often lead to undesirable outcomes in the presence of learning, resulting in overly greedy or aggressive behavior \cite{leibo2017multi,lerer2017maintaining}. Humans on the other hand display the ability to overcome these pathologies through the evolution of social structures (also called institutions) \cite{ostrom1990governing,janssen2008turfs}, which renders the question of how known structures can be utilized or learned in the context of multi-agent systems.

\begin{figure}[hbtp]
\centering
\includegraphics[width=0.31 \textwidth]{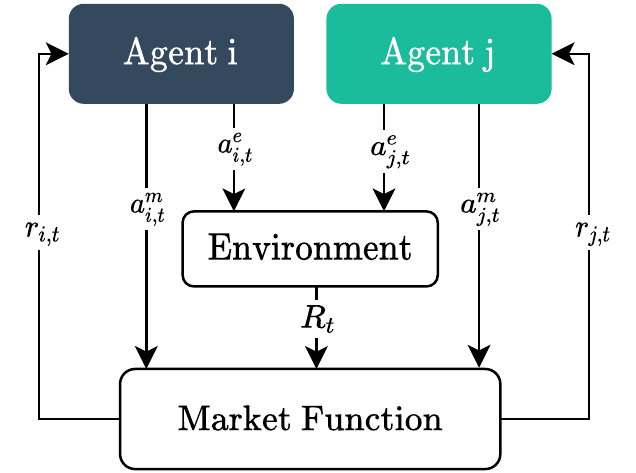}
\caption{Stochastic Market Game: In addition to their environmental actions at time $t$, agents choose market actions, which are processed by a market function to compute agent specific rewards.}
\label{fig:SMG}
\end{figure}

Effective institutions shape the interaction of individuals, by setting constraints or providing specific incentives to the members of a society. Markets for instance are effective means to organize large numbers of self-interested entities in a decentralized manner, with a rich body of literature on markets dating back to Adam Smith \cite{smith1937wealth}. Besides its attributes of openness and scalability, as \cite{miller1988markets} note, a market is particularly effective in promoting efficient, cooperative interaction between heterogeneous participants because (idealized) markets define distinctive rules (property rights, unforgeable currencies, trademarks), such that being cooperative becomes a dominant strategy in the game theoretic sense. While markets have been suggested to coordinate distributed decision making in fully cooperative multi-agent systems \cite{miller1988markets,holland1985properties,chang2020decentralized}, in this work we adopt markets to change the dynamics of a given mixed-motive game to promote cooperation and make defective strategies less attractive.



Our main argument is that modelling agents as market participants turns agents' self-interest from a flaw into a feature, as voluntary trades are Pareto improvements (in the absence of external costs) and hence direct the overall outcome towards the Pareto optimum of the solution space. We introduce markets for a given multi-agent problem by extending the formulation of Stochastic Games (or Markov Games) \cite{littman1994markov} with 1): a set of dedicated market actions that allow agents to become active in a given market, and 2): defining a function that redistributes rewards based on agents' market behavior. We term this a \textit{Stochastic Market Game}.





\section{Related Work}\label{sec:related-work}
\paragraph{Markets in cooperative games.}
First proposals for utilizing market forces to decentralize computationally complex tasks date back to \cite{miller1988markets,miller1988comparative}, with probably the first algorithm for cooperative multi-agent systems given in \cite{holland1985properties}, called the Bucket Brigade. The algorithm works on the basis of entities called classifiers, which are composed of a condition and an action. When the system state matches the condition of an entity, it computes a bid and the entity with the highest bid is allowed to push its specific action onto an execution queue. The bid is distributed among the previously active classifiers that led the system to the triggering condition, such that chains of rewards originate and learning may take place. \textit{Hayek machines} \cite{baum1999toward} extend the original Bucket Brigade by property rights (to act in the world) through agents which can invest in their own descendants, in order to profit from their offspring later. Market-based reinforcement learning (RL) as a means to learn in partially observable Markov Decision Processes (POMDPs) has been suggested in \cite{kwee2001market}, which is based on a variant of the Hayek machine with an added memory register that can be manipulated by agents with specific write actions. More recently, the idea of buying and selling the right to act in the world by individual agents has been taken up in \cite{chang2020decentralized} and combined with different RL algorithms to formalize a general framework for organizing a society of agents. In contrast to the approaches described above, the focus of this work lies in non-cooperative or mixed settings, where agents have individual goals. We therefore do not seek to solve the credit assignment problem with market mechanisms but suggest markets as a catalyst to allocate resources to agents in order to improve overall efficiency.



\paragraph{Non-cooperative systems.}
In \cite{perolat2017multi}, multi-agent reinforcement learning is used to analyze the emergent equilibria for common-pool resource appropriation settings, for which game theoretic models predict the tragedy of the commons. In \cite{leibo2017multi}, sequential social dilemmas (SSDs) are introduced, which extend social dilemmas of game theory to better capture real world aspects, such as temporal extension and non-binary grades of cooperation. The idea of trading actions in SSDs for independent agents has been proposed in \cite{schmid2018action}, where it has been shown that the approach promotes cooperative strategies. However, the empirical evaluation has been limited to two agents. In this work, Stochastic Market Games are proposed, providing a general framework for agents to learn the distribution of rewards as a meta game. In this work, the action market is categorized as an instance of a conditional market and reevaluated with up to $16$ agents. More recently, \cite{lupu2020gifting} proposed gifting, a mechanism that allows agents to directly assign reward to other agents, used as a means to overcome the tragedy of the commons. The gifting mechanism is similar to the market approach, as it extends action spaces with designated gifting actions. However, it differs from the idea of a market, as there is no actual trading between agents based on a buying and selling offer.

market-games\section{Background}

In this work we consider systems comprised of multiple independent agents formalized as a Stochastic Game $\mathcal{G}$ (also called Markov game) \cite{littman1994markov}, which is defined by a set of states $\mathcal S$, and a collection of action sets $\mathcal A_1,...,\mathcal A_N$, one for each agent in the environment. Given a current state and one action from each agent the next state of the environment is defined by the transition function $\mathcal P : \mathcal S  \times \mathcal A_1 \times...\times \mathcal A_N \to PD(\mathcal S)$ where $PD(\mathcal S)$ is the set of discrete probability distributions over $\mathcal S$. Each agent has a reward function $ r_i : \mathcal S \times \mathcal A_1 \times...\times \mathcal A_N \to \mathbb R$, that defines its goal. An agent tries to learn an optimal policy $\pi^{i}$ that maximizes its expected sum of discounted rewards (called return) $\mathcal R_{i,t} := \sum^{\infty}_{t=1} \gamma^{t-1}r_{i,t}$ with discount factor $\gamma < 1$ \cite{sutton2018reinforcement}. In this work we consider scenarios of mixed motives, which means agents' rewards are correlated to varying extents so depending on the situation their goals might be aligned or be in conflict \cite{dafoe2020open}. 

To model learning, we utilize independent learning, a decentralized reinforcement learning (RL) approach where agents independently follow individual learning rules such as Q-learning or policy gradient methods. With \emph{deep} RL each agent's policy is represented as a deep neural network denoted $\pi^{i}_{\theta}$ where $\theta$ are the trainable parameters of a neural network. Independent learning is popular in mixed motive scenarios, as no assumptions on each agent's goal are required during training. One drawback of this this paradigm is that agents implicitly model other agents as being part of the environment so in general IL suffers from the issue of non-stationarity (agents have a moving target). In practice non-stationarity can be mitigated by letting agents adapt to a slowly changing data distribution which stabilizes learning \cite{leibo2017multi}. 

\paragraph{Independent Deep-Q-Networks (IDQN).} IDQN models each agent $i$ as a an instance of a \textit{Deep-Q-Network (DQN)} \cite{mnih2015human}, gathering its own experience represented as tuple $(s, a, r, s')$ comprising a state $s$, an action $a$, a reward $r$ and a next state $s'$. At each training step a minibatch of transitions is sampled to update the state-action value function  $Q_{i} : \mathcal{S} \times \mathcal{A}_{i} \to \mathbb R $ according to the update rule: $Q_{i}(s, a) \leftarrow Q_{i}(s, a) + \alpha \big[ r_{i} + \gamma \max_{a' \in \mathcal A^{i}}  Q_{i}(s', a') - Q_{i}(s,a) \big]$ where $\alpha$ is the learning rate. From $Q_i$ a policy is derived by selecting actions according to their $Q$-values in a given state or (to encourage exploration) randomly with probability $\epsilon$.


\paragraph{Independent Actor-Critic (IAC).}
With IAC each agent is modelled as an instance of an actor-critic algorithm. Actor-critic algorithms seek to learn optimal policies by combining value function learning with policy gradient based methods. The difference of policy based methods and value based methods is, that policy based methods directly search the policy space for the optimal policy $\pi^{*}_{\theta}(a|s)$ with parameters $\theta$. Here we utilize Proximal Policy Optimization (PPO)~\cite{schulman1707proximal}, an algorithm that ensures policy updates are relatively small to mitigate the negative effects associated with drastic changes in the policy. PPO achieves this by keeping the probability ratio
$r_t(\theta) = \frac{\pi_{\theta}(a_t|s_t)}{\pi_{\theta_{old}}(a_t|s_t)}$ close to $1$ with a clipped surrogate loss.



\section{Stochastic Market Games}\label{sec:approach}
In situations of mixed-motives, in which agents' interests are possibly conflicting, a central hurdle towards cooperation is that agents do not know how their behavior affects others. Specifically, agents neither receive any feedback on how their actions affect payoffs of fellow agents, nor on costs that they might induce to others. These factors are therefore external from the perspective of individual agents, as they are not considered in their individual reward functions. Moreover, there is no incentive for individual agents to internalize external costs as long as they are not being compensated or receive something in return. Combined, these aspects can diminish the overall reward as under such circumstances individual rational decision making will not necessarily lead to Pareto optimal results \cite{leibo2017multi}.

In this work, we propose to enrich mixed-motive scenarios with a market mechanism, to change the dynamics of the game through rules that generate incentives so as to turn cooperation into a dominant strategy. We argue that by entering mutual agreements via trades, returns of independent agents become interdependent, which helps to internalize former external costs (and profits).  We call a Stochastic Game extended by a market a Stochastic Market Game (SMG), which we formally define as a tuple: $(\mathcal{G}, \mathcal{A}_1^{m},...,\mathcal{A}_N^{m}, \mathcal M)$ consisting of 
\begin{itemize}
    \item The underlying Stochastic Game $\mathcal{G}$ with a set of states $S$, a set of environmental actions $\mathcal{A}_1^{e},...,\mathcal{A}_N^{e}$, the transition function $\mathcal P: S \times \mathcal{A}_1^{e} \times... \times \mathcal{A}_N^{e} \to PD(\mathcal S)$, and reward functions $r_{i}: \mathcal S \times \mathcal{A}_1^{e} \times...\times \mathcal{A}_N^{e} \to \mathbb{R}$ for each agent $i$.
    \item A collection of market action sets $\mathcal{A}_1^{m},...,\mathcal{A}_N^{m}$, one for each agent. 
    \item A market function $\mathcal M : \mathcal S \times \mathcal{A}_1^{m} \times...\times \mathcal{A}_N^{m} \times \mathbb{R}^{N} \to \mathbb{R}^{N}$, that computes the redistribution of rewards based on the state, agents' market actions, and rewards.
 \end{itemize}
 
The Stochastic Market Game allows agents to become market participants by applying market actions $a^{m} \in \mathcal{A}^{m}_{i}$ in addition to their environmental actions $a^{e} \in \mathcal{A}^{e}_{i}$ (see Figure \ref{fig:SMG}). We differentiate two types of market functions $\mathcal M$, which we call \textbf{unconditional markets} and \textbf{conditional markets}. An unconditional market allows trades between two agents that are not constrained through further conditions. Therefore, agents connected via an unconditional trade need not commit to specific behavior, rules or norms, hence unconditional trading is a way to interconnect payoffs without restricting agents' individual behavior. In contrast, a conditional market defines a trade as a bilateral commitment between agent $i$ and agent $j$ to make a transaction that is triggered when a specified condition occurs, such as specific events, states, or certain behavior of fellow agents.  We build our definition of a market on the concept of idealized markets \cite{miller1988comparative} that define rules such that individual participants cannot violate property rights, deceive or cheat other agents, as these laws are enforced by the environment.

\begin{figure}[hbtp]
\centering
\includegraphics[width=0.34\textwidth]{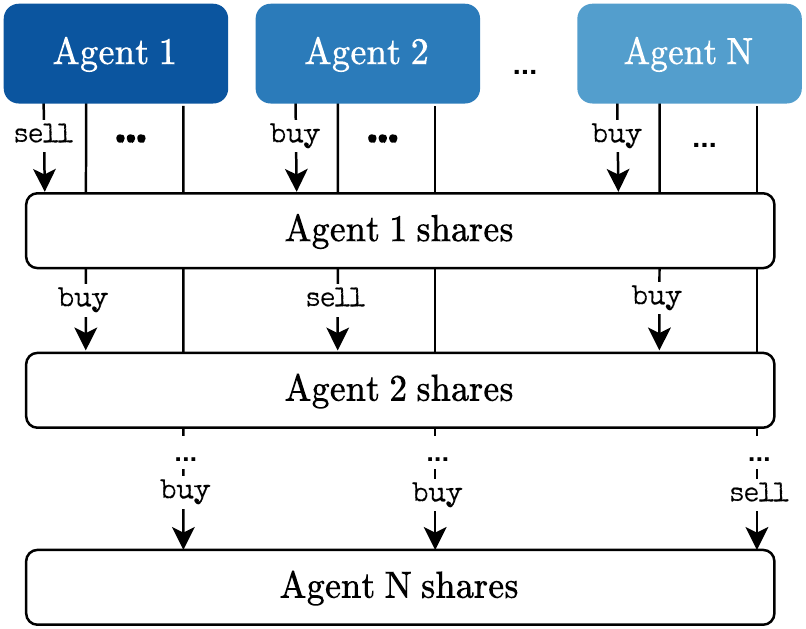}
\caption{Shareholder market: agents buy shares to participate in other agents' returns or sell shares to let others participate in their own return.}
\label{fig:shareholder-market}
\end{figure}

\subsection{Unconditional Markets}
Unconditional markets are inspired by the idea that in order to align agents' goals, they need to be enabled to let other agents participate in their own return. This is similar to a company, issuing shares from which shareholders earn a dividend. In such an agreement the shareholder is not required to fulfill any specific actions in order to receive the dividend, hence the market works unconditionally. Transactions in such a market require agents to decide on whether to buy shares from other agents and whether to sell their own shares to others at each step. In this work we will use the terms unconditional market and shareholder market (SM) synonymously. 

Formally in the shareholder market agent $i$'s market action $a^{m}_{i,t}$ at time $t$ specifies the willingness to sell to agent $j$ or buy from agent $j$ respectively (see Figure \ref{fig:shareholder-market}). If the buying request from $i$ coincides with a selling offer from $j$ (or vice-versa), the transaction is done and $i$ holds a share from $j$. From there on, the share lets $i$ participate in $j$'s return by a fixed dividend $d$. Optionally, shares can also have a price $p$ that must be paid for buying the share in the first place. However, such a price is not mandatory, since agents with high stakes will be willing to give their shares freely to make other agents shareholders of them to align their goals.

\subsection{Conditional Markets}
The conditional market is based on the idea of allowing agent $i$ to transfer reward to agent $j$ in direct response to $j$'s behavior. In this sense, a conditional market defines a contract of sale between a seller and a buyer side, similar to real world contracts where two persons agree on a payout when the contract conditions are met. Here we consider the case where contract conditions can be specified via the market actions $A_{i}^{m}$, and can be based either on the current state $s_{t}$ or on the environmental actions $a^{e} \in \mathcal{A}^{e}_{i}$ that are executed by agents. Again we define the environment as a trusted entity that verifies the specified conditions, so agents are prevented from cheating or breaking a contract. 

An instance of a conditional market where the conditions are based on agents' actions has been proposed by \cite{schmid2018action}, as a so called action market (AM). In an action market, agent $i$'s action $\Vec{a}_{i,t}$ at time $t$ comprises both, an environmental action $a_{i,t}^{\texttt{env}}$ and possibly an offer action $a_{i, j,t}^{\texttt{offer}}$ towards agent $j$. If $j$ happens to execute the suggested action at time $t$, i.e. if  $a_{j,t}^{\texttt{env}}$ equals $a_{i,j,t}^{\texttt{offer}}$, then a trade between $i$ and $j$ is established. In this case, $i$ is obliged to give a predetermined price $p$ in terms of reward to $j$.

\subsection{Learning Market Games}
We now describe the training procedure for $N$ independent learners combined with our market approach. Here, the training is defined for episodic tasks, where learning is done over the course of $E$ episodes, each comprising a maximum number of $T$ steps (pseudocode  given in Algorithm \ref{alg:ILM}). For the training with a market, the market function $\mathcal M$ is initialized at the beginning of the training and called after each environmental step. The market function $\mathcal M$ returns the adjusted rewards with respect to the trades between agents.

The principle of the market function is shown in Algorithm \ref{alg:market-function}. Agents can have either claims or liabilities towards other agents, which are accumulated during the course of an episode and recorded in a balance sheet $b_t \in \mathbb{R}^{N \times N} \subset s_t$. At each market step, the actual trades are computed on the basis of agent's buying and selling offers. These offers are stored in the selling matrix $S \in N\times N$ and the buying matrix $B \in N \times N$ respectively. The distinction between the action market and the shareholder market lies in the way buying and selling offers are defined. In the action market, a selling offer $a_{i,j}^{\texttt{sell}}$ from agent $i$ towards agent $j$ is represented by $i$'s environmental action, whereas a buying offer $a_{i,j}^{\texttt{buy}}$ from $i$ to $j$ is defined as the environmental action that $i$ is willing to buy from $j$, hence $a_{i,j}^{\texttt{buy}}, a_{i,j}^{\texttt{sell}} \in A^{e}$. In the shareholder market, the selling and buying offer do not specify terms for the transaction (unconditional), therefore represent a binary choice: $a_{i,j}^{\texttt{buy}}, a_{i,j}^{\texttt{sell}} \in \{0, 1\}$.

In case agents can trade with all other agents at a given time step, the action space increases exponentially with the number of agents, since agents need to decide for all other agents whether they want to make a buying or a selling offer. For the action market, the increase is even larger than for the shareholder market, as all offers are conditioned on an action which increases the number of possible offers by the number of available environmental actions. However, in practice action spaces can be effectively limited, by either limiting the number of possible offers at a time or by applying offers to groups of agents. In this work, we allow agents to make a buying offer only towards a single agent in the action market, through which the increase in action space is only linear in the number of agents: $|\mathcal A| = |\mathcal A^{e}| + ((N-1) *\mathcal |A^{e}| * |A^{e}|)$. For the shareholder market we use a variant that we term broadcasting offers, which means that the selling offer has no designated recipient so an agent willing to sell accepts all offers at a each step. 

\begin{algorithm}
\caption{Independent Learning with Market}\label{alg:ILM}
  \begin{algorithmic}[1]
  \For{\textit{episode $e$ $\in$ $E$}}
        \For{\textit{step $t$ $\in$ $T$}}
            \For{\textit{agent ${i}$ $\in$ $N$}}
                \State $\Vec{a}_{i,t} \sim \pi^{i}(s_{t}|\theta^{i})$\Comment{sample action}
            \EndFor
            \State $s_{t+1} \gets \mathcal P(\Vec{a}_{t}|s_t)$\Comment{episode step}
            \State $\Vec{r}_{t} \gets \mathcal \mathcal R(\Vec{a}_{t}|s_t)$\Comment{get rewards}
            \State $\Vec{r}_{t},  \gets \mathcal M(\Vec{a_{t}}, \Vec{r}_{t}, s_t)$\Comment{call market} 
            \For{\textit{agent ${i}$ $\in$ $N$}}
                \State train $\theta^{i}$ based on $s_{t}, \Vec{a}_{i,t}, s_{t+1}, \Vec{r}_{i,t}$
            \EndFor
        \EndFor
    \EndFor
\end{algorithmic}
\end{algorithm}

\begin{algorithm}
  \caption{Market Function}\label{alg:market-function}
  \begin{algorithmic}[1]
\Procedure{$\mathcal M(s_t, \Vec{a}_{t}, \Vec{r}_{t})$}{}
    \State $b_t \gets s_t$\Comment{get balance from current state}
    \State $S, B \gets 0_{N, N}$
    \For{\textit{agent ${i}$ in $N$}}
        \For{\textit{agent ${j}$ in $N \setminus i$ }} 
            \State $S_{i,j} \gets a_{i,j}^{\texttt{sell}}$\Comment{register sell}
            \State $B_{i,j} \gets a_{i,j}^{\texttt{buy}}$\Comment{register buy}
        \EndFor        
    \EndFor
    \State $T \gets S \wedge B$\Comment{compute trades}
    \State $\Vec{r}_{t} \gets$ set rewards w.r.t. to $T$, $b_t$ and $\Vec{r_t}$
    \State \Return $\Vec{r}_{t}$
\EndProcedure
\end{algorithmic}
\end{algorithm}



Moreover, agents can only make either a  single selling or a single buying offer at a time, which reduces the size of the action space to: $|\mathcal A| =  |\mathcal A^{e}| * |\{0, 1\}| *(N)$. The trades at time $t$ are computed by the logical conjunct of the selling offers with the buying offers. Afterwards, the rewards are computed with respect to the current trades and agents' accumulated paying obligations from earlier trades, that have not yet been settled. An agent is solvent, i.e. able to pay its liabilities, in case it has earned a positive reward. Otherwise, agents have a liability to another agent, which is settled whenever the agent becomes solvent. Note that there can arise situations where paying obligations cannot be settled during the course on an episode. Optionally, agents can be allowed to make debts by paying their liabilities even if they received a non-positive reward, i.e., $r_{i,t} \leq 0$.

\section{Experiments}\label{sec:evaluation}
In this section we evaluate the proposed market mechanisms in different two player matrix games before we extend the evaluation towards temporally and spatially extended domains where we evaluate with up to $16$ independent agents.

\subsection{Two Player Games} 
\begin{figure}
\centering
     \subfloat[Conflict game \label{fig:matrix-game}]{
       \includegraphics[align=c,width=0.17\textwidth]{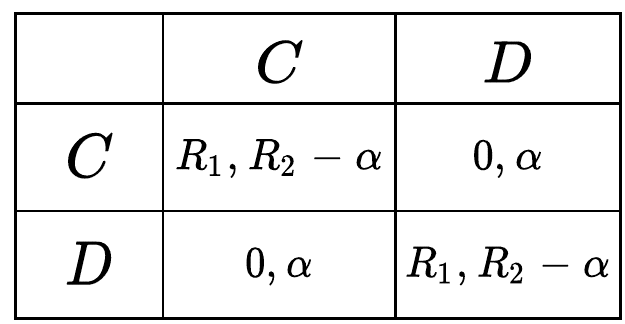}
     }
     \hfill
     \subfloat[Overall reward \label{fig:rewards-matrix-game}]{%
       \includegraphics[align=c,width=0.29\textwidth]{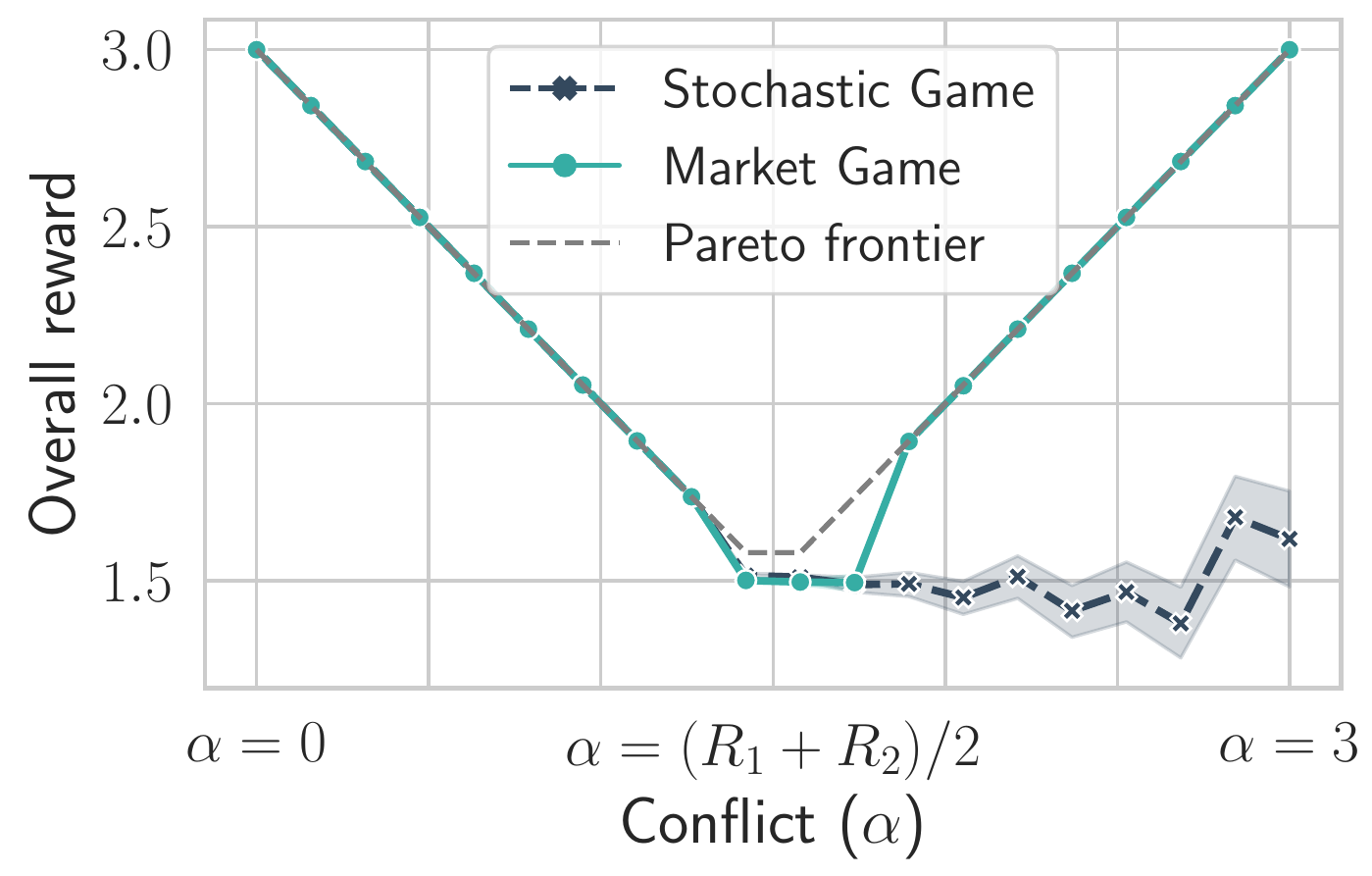}
     }
     \caption{Independent Q-learning in the Conflict game for varying conflict $\alpha$.}
     \label{fig:market-dynamics}
\end{figure}

Our first analysis aims at answering the question of how the level of conflict between two players affects the overall outcome and how the proposed market can be utilized in order to resolve the conflict in such a scenario. To that end, we propose the two player \textbf{Conflict game}, depicted in Figure \ref{fig:matrix-game}, where each player faces the choice between two actions $C$ and $D$. The joint action (vector of actions from both players) $(i, j)$ with $i, j \in \{C, D\}$ yields a reward of $(x, y)$ as given in the respective cell of the matrix, where $x$ is the reward for player $1$ (column player) and $y$ is the reward for player $2$ (row player). The amount of conflict between the players in this game can be controlled by the parameter $\alpha$: For $\alpha = 0$ and with $R_1, R_2 > 0$, the players' goals are perfectly aligned, so they only face a coordination challenge, which means they need to choose the same action in order to receive a positive reward. However, for growing values of $\alpha$ the players' goals start to diverge as the column player increasingly becomes incentivized to prefer an uncoordinated outcome, i.e. when player $1$ chooses $C$ then player $2$ prefers $D$ and vice versa. Finally, for $\alpha > R_2$ their goals are completely opposing each other so each player will only receive a positive reward if the other player fails to do so, in which case there is no Nash-equilibrium in pure strategies.

For the experiment we set $R_1 = 0.375$ and $R_2 = 2.625$, so $R_1 + R_2 = 3$. Note however, that depending on $\alpha$ the maximum achievable overall reward can be smaller than $3$, since due to the conflict between the players some loss of reward is unavoidable. We therefore indicate the maximum achievable reward corresponding to a specific value of $\alpha$ with a dashed line in Figure \ref{fig:rewards-matrix-game} (called Pareto frontier), which has a V-shape: for $0 < \alpha < 2 / (R_1 + R_2)$, the maximum overall reward decreases before at $\alpha = R_1 + R_2 / 2$ it starts to increase again, as at this point agent $2$ will get more reward in case of choosing action $(C, D)$ or $(D, C)$ than both players can get from choosing $(C, C)$ or $(D, D)$.

We now analyze the learning dynamics with and without market between two independent learners in this game. To that end two independent Q-learners were trained over $25$ independent runs for each level of conflict $\alpha$, where each run consisted of $4000$ consecutive steps (see Figure \ref{fig:rewards-matrix-game}). With standard Q-learning and in case of no market, the overall reward from both agents follows the shape of the Pareto frontier up to the point where the Pareto frontier increases again. At this point, the increasing conflict between the two players cannot be resolved by the agents since their goals diverge. This loss in overall reward can be mitigated by introducing the action market mechanism in the following way: agent's available action spaces are extended with additional actions to allow the transfer of reward, thus the action space is extended from $\{C, D\}$ to $\{(C,-), (C ,C), (C,D), (D,-), (D,C), (D,D)\}$, where the first component indicates the player's own action and the second component is the offer towards the other player ($(C,-), (D, -)$ refer to actions with no intended trade). As is shown in Figure \ref{fig:rewards-matrix-game}, the overall reward achieved with the same Q-learning configuration in the presence of a market is identical or close to the Pareto frontier for all levels of $\alpha$, which demonstrates how the market serves as a catalyst to align agents' increasingly diverging goals.



\begin{figure}
\centering
     \subfloat[Prisoner's Dilemma \label{fig:pd-env}]{
       \includegraphics[align=c,width=0.17\textwidth]{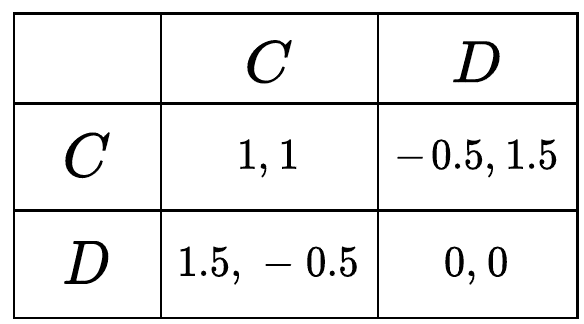}
     }
     \hfill
     \subfloat[Overall reward \label{fig:rewards-pd}]{%
       \includegraphics[align=c,width=0.29\textwidth]{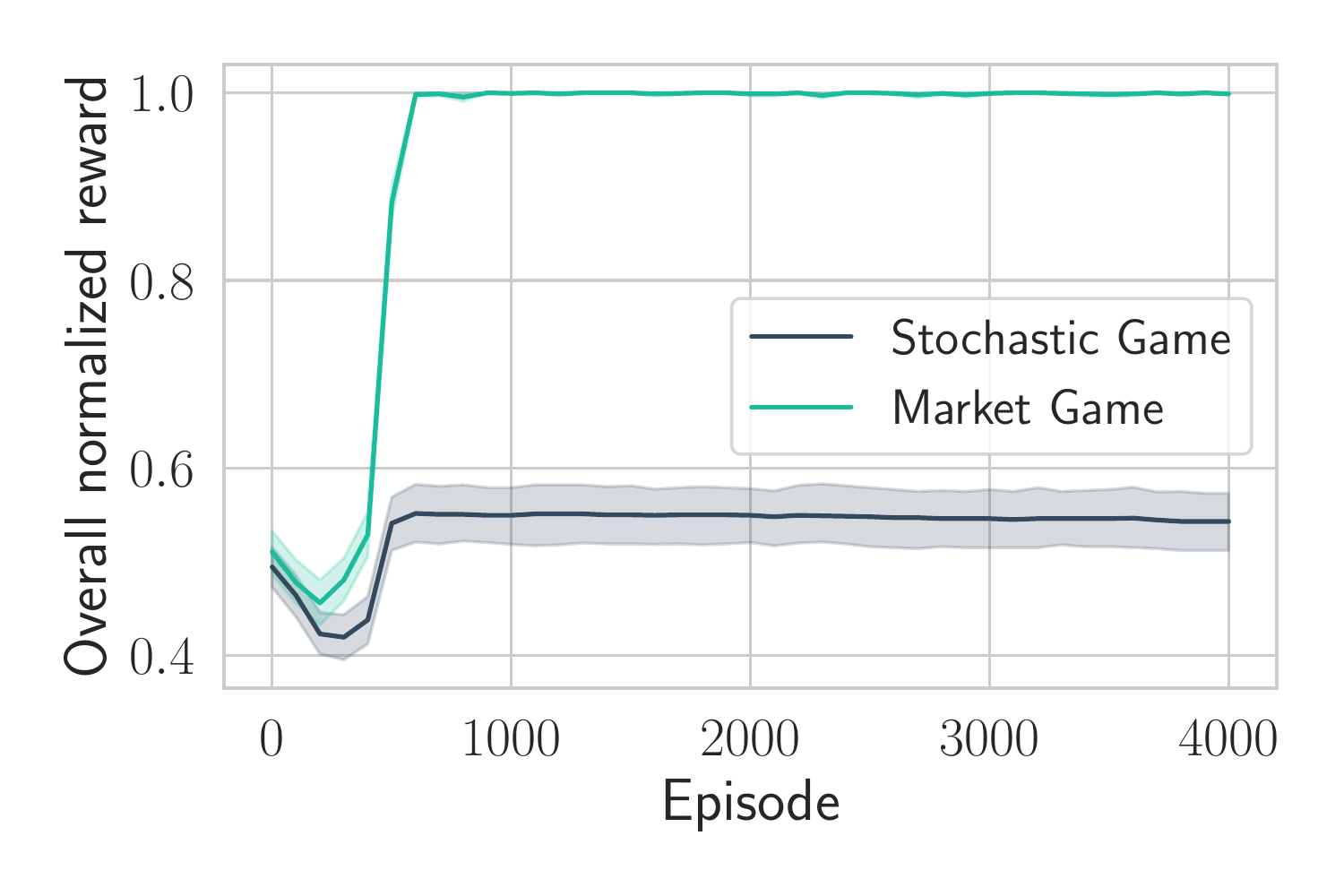}
     }
     \caption{Comparison of independent Q-learning in the Prisoner's Dilemma with and without market.}
     \label{fig:PD}
\end{figure}

We now turn to a two player scenario, called the Iterated \textbf{Prisoner's Dilemma} shown in Figure \ref{fig:pd-env}, which poses a challenge to cooperative behavior, since in a one-shot game, where the game is only played once a rational player will choose the globally non-optimal action $D$ (defection) even though both players can be better off when they both choose to cooperate (denoted $C$). In an iterated version of the Prisoner's Dilemma cooperation might emerge based on reciprocity. However, standard reinforcement learning is unlikely to produce meaningful cooperation based on reciprocity, since it requires recursive reasoning about the consequences of one's own behavior on others, which is not part of model-free RL. This circumstance can be seen in Figure \ref{fig:rewards-pd}, where we show results from Q-learning for $4000$ consecutive steps of the iterated Prisoner's Dilemma averaged over $1000$ independent runs with returns normalized between $0$ and $1$. As is shown, independent Q-learning is likely to converge to the unique, non-optimal Nash-Equilibrium ($D,D$), indicated by the low level of overall reward. Through the extension with a market in the same way as before, the learned strategies by independent Q-learning can be improved towards optimal play, that is cooperative play in all games as is shown in Figure \ref{fig:rewards-pd}. Interestingly, for the iterated Prisoner's dilemma we noted that only a negative price helps to increase cooperation (here we used $p=-1.9$) in which case agents' strategies are consistently changed to mutual cooperation $(C, C)$ after approximately $500$ steps, thereby achieving the maximum reward.

\begin{figure}
     \subfloat[Return 8 agents $\textit{SF}$ \label{fig:rewards-8-agents-sm}]{
       \includegraphics[width=0.24\textwidth]{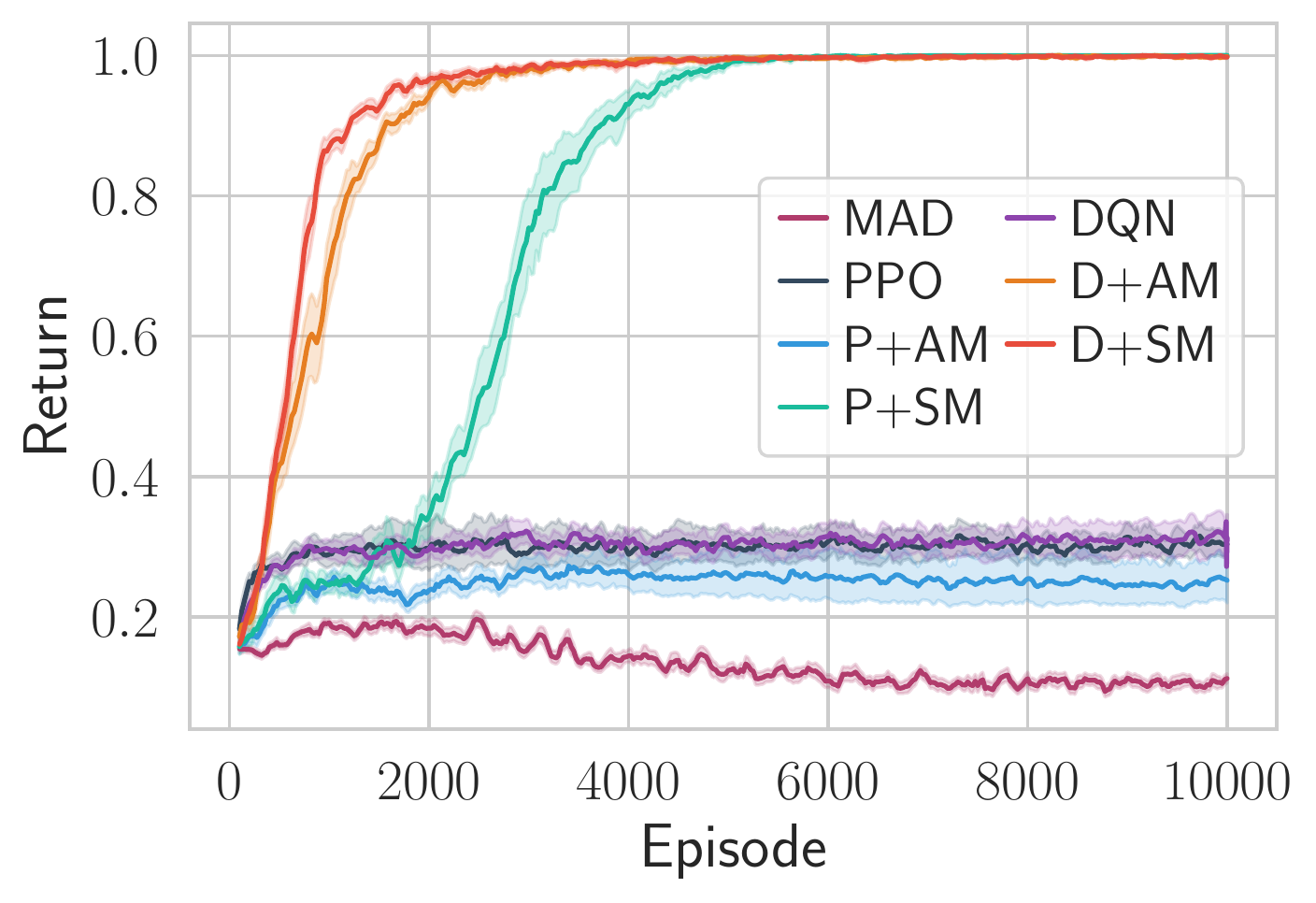}
     }
     \subfloat[Return 16 agents $\textit{SF}$ \label{fig:rewards-16-agents-sm}]{%
       \includegraphics[width=0.24\textwidth]{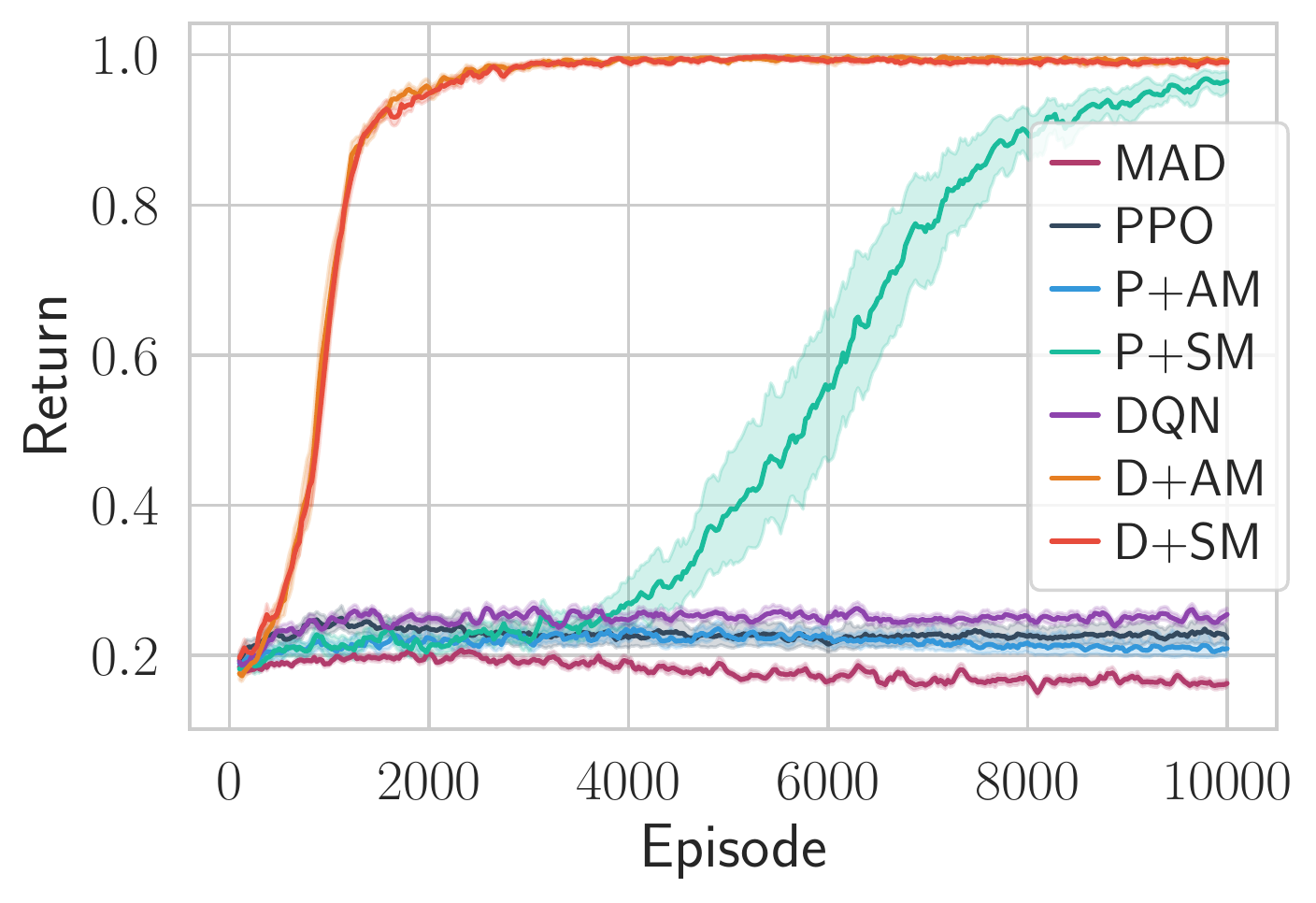}
     }
    \hfill
     \subfloat[Return 8 agents $\textit{RF}$ \label{fig:rewards-8-agents-r}]{%
       \includegraphics[width=0.24\textwidth]{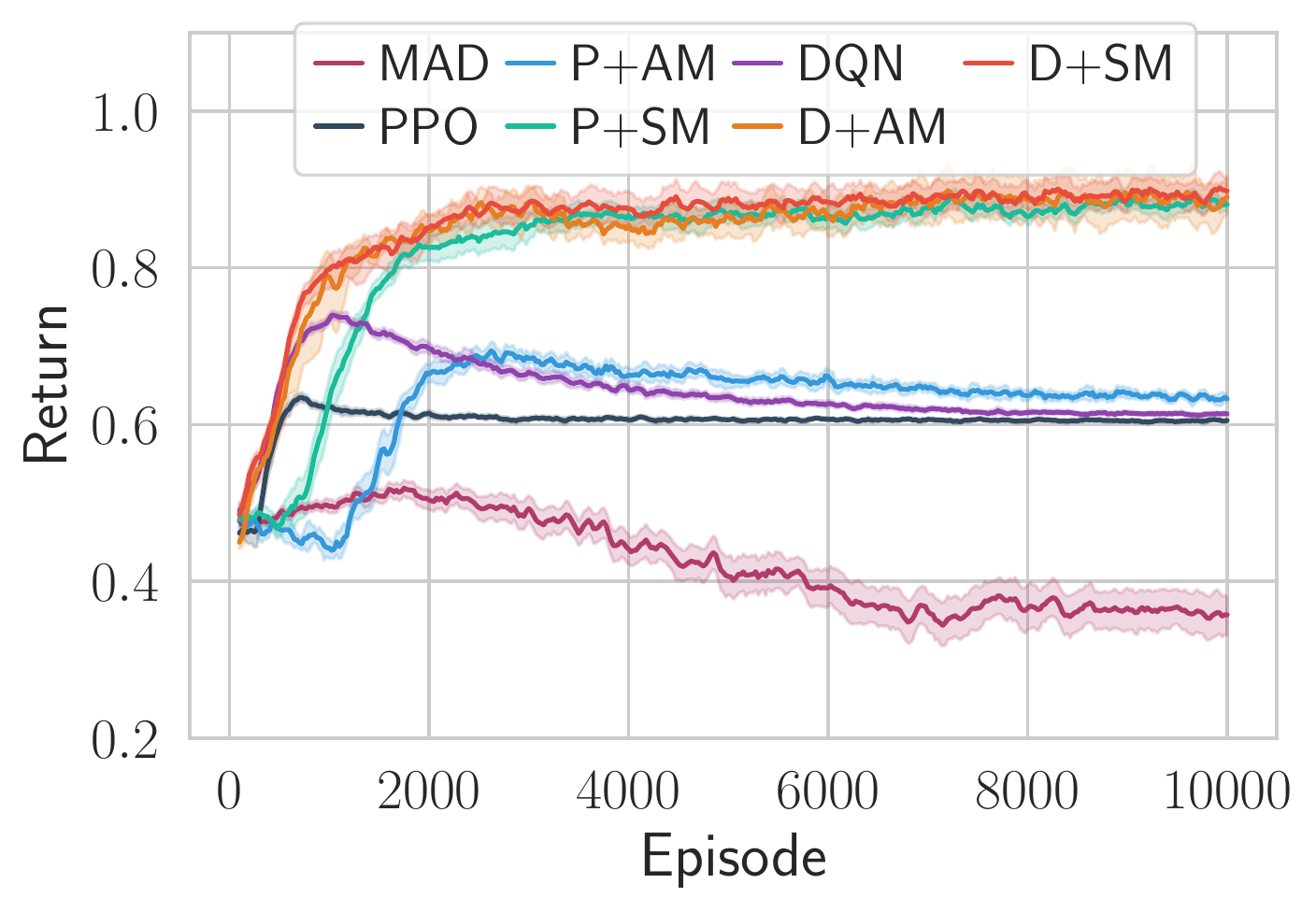}
     }
     \subfloat[Return 16 agents $\textit{RF}$ \label{fig:rewards-16-agents-r}]{%
       \includegraphics[width=0.24\textwidth]{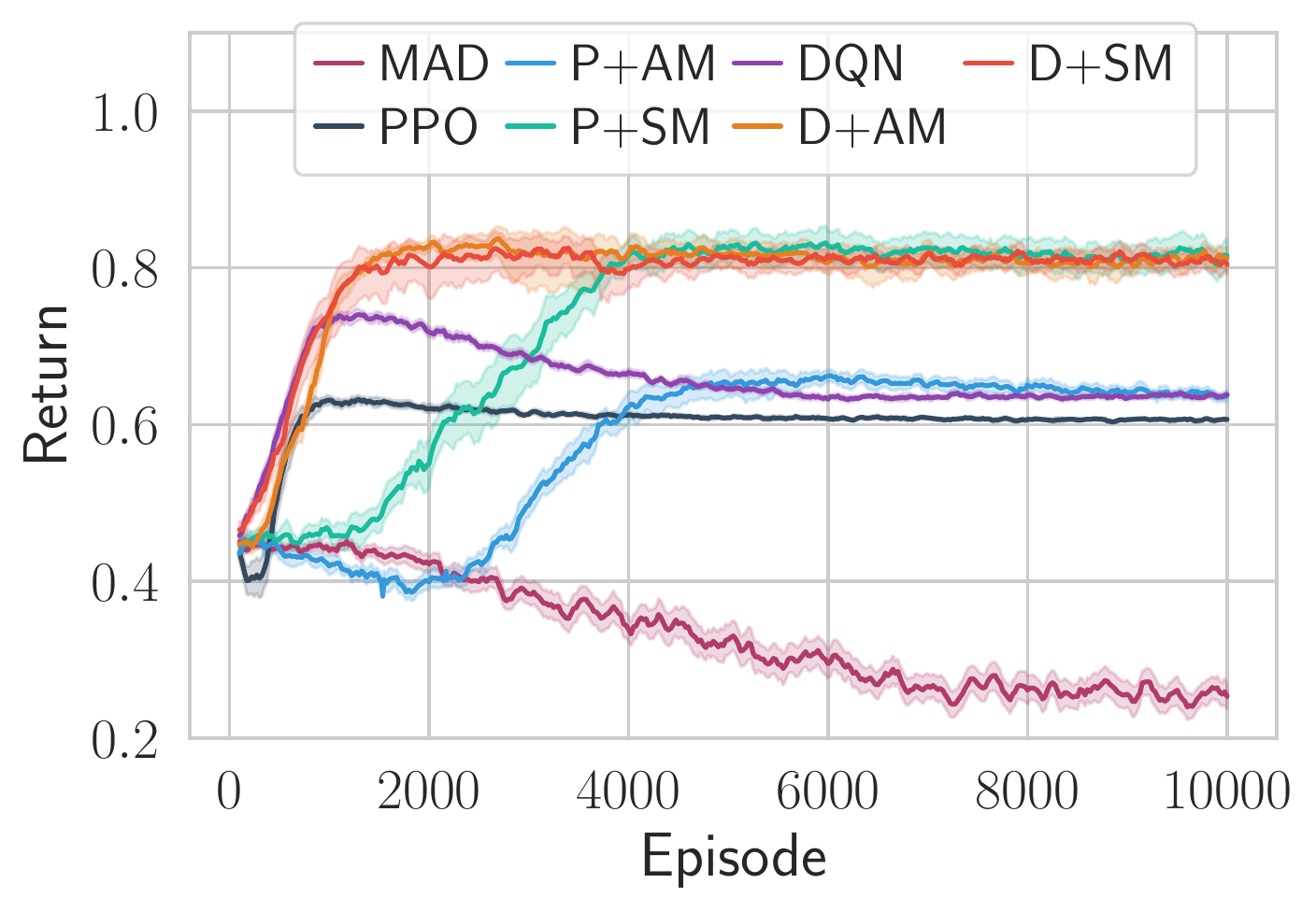}
     }
     \caption{Normalized return from $20$ independent runs of the Smartfactory ($\textit{SF}$, a and b) and Refinery ($\textit{RF}$, c and d) for MADDPG (MAD), PPO and DQN without market and with either shareholder market (P+SM, D+SM) or action market (P+AM, D+AM).}
     \label{fig:rewards}
\end{figure}

\subsection{N-Player Games}
We now turn to more complex scenarios, which scale in the number of agents and are inspired by resource allocation problems, where agents follow individual goals but share limited resources (e.g. machines). The first domain, called \textbf{Smartfactory} (\textit{SF}), holds $m$ different machine types that are to be used by the agents. For every episode, each agent receives a task that is comprised of $n$ machines, i.e. an agent must activate all $n$ machines (by visiting them) before it is considered to be done. Rewards are sparse, i.e. they are only given to an agent whenever it has visited all machines that are part of its task. There are two parameters to fine-tune the potential conflict between agents: $t_{\text{inactive}}$ regulates the number of time steps that a machine is inactive after usage (and thus not usable by any other agent). Consequently, competition between agents increases with increasing $t_{\text{inactive}}$. The second parameter concerns the urgency or priority of an agent to fulfil its task, with two possible priorities (high or low). If an agent’s task is assigned high priority, that agent receives a higher reward $r_{\textit{high}}$ after completion than the agent assigned with the low-priority task (which receives a reward defined by $r_\textit{low}$). An increasing gap between $r_{\textit{high}}$ and $r_{\textit{low}}$ increases the potential benefit of coordination and cooperation between agents in terms of overall reward.

The second domain, called \textbf{Refinery} is inhabited by two types of agents, namely \emph{refining} and \emph{consuming} agents, which compete for a common resource. Furthermore, the resource appears in two different classes, i.e. unrefined and refined. The distinguishing feature of a refining agent and a consuming agent is that the refiner has the choice of either consuming the unrefined resource, which yields a relatively small reward $r^{\textit{low}}$, or to refine the resource, in which case it can no longer be consumed by himself. A consuming agent, on the other hand, cannot refine the raw resource but can only consume the already refined units, which gives a high reward $r^{\textit{high}}$ to the consuming agent. The dilemma in this scenario is that the overall return is maximal if all available resources are refined and subsequently consumed. However, as a refining agent cannot consume the refined good, it is individually rational to consume the raw resource to achieve at least a small reward. In all runs we set $r_{\textit{high}} = 5, r_{\textit{low}} = 1$  $t_{\text{inactive}} = 8$ in the Smartfactory and $r_{\textit{high}} = 5, r_{\textit{low}} = 0.02$ in the Refinery.

For the evaluation, we compare the settings: DQN and PPO without markets (Stochastic Game baseline), DQN and PPO with action market (AM) or shareholder market (SM). We also show results from the Multi-Agent Deep Deterministic Policy Gradient (MADDPG) algorithm \cite{lowe2017multi} as a baseline featuring a centralized critic. All settings were tested with $4$, $8$ and $16$ agents with the overall return from $8$ and $16$ agents displayed in Figure \ref{fig:rewards} (normalized overall return, averaged over $20$ independent runs). 

\begin{figure}[hbtp]
\centering
\includegraphics[width=0.34 \textwidth]{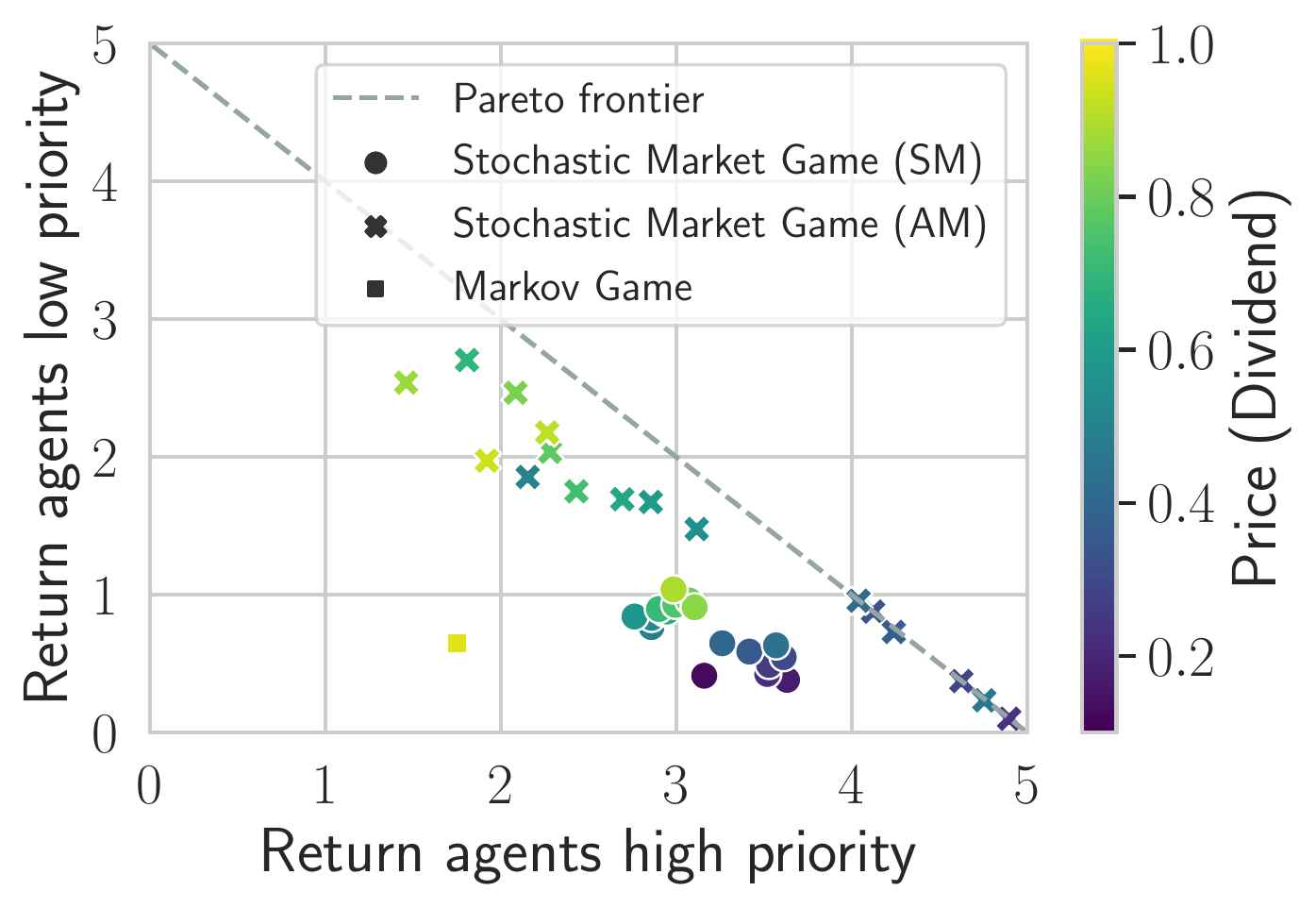}
\caption{Market reward distribution for different price levels.}
\label{fig:distribution}
\end{figure}

In all settings, the highest reward is achieved when either the action market or shareholder market is available. Learning without markets cannot achieve any outcome that goes beyond the expected equilibrium outcome under self-interested optimization. We presume one reason for the difference in outcomes between DQN and PPO in combination with a market is due to the difference in the size of the action spaces, where it seems that DQN can cope more easily with the increased action space in both market forms. This is also underlined by the drop of performance of PPO combined with an action market, which has a comparatively larger action space than the shareholder market.




To further analyze outcomes with respect to the distribution among agents and their overall efficiency, we visualize the distribution of the returns, grouped for agents according to their task priority and for different levels of prices (dividends) in Figure \ref{fig:distribution}. Each point $(x, y)$ comprises the share of the overall reward from agents with high priority $x$, and the reward share of agents with low priority $y$. For all tested prices, the outcomes are closer to the Pareto frontier (more efficient) in the presence of a market. Interestingly, the efficiency of outcomes decreases for higher price levels in case of a shareholder market, whereas the results are less affected by the price in the action market, which does not influence the division of the return as much as the price levels in a shareholder market. Finally, the dividend and price level influences the division of the return between agents with high and low priority tasks, such that higher prices lead to an increased share of the return being assigned to agents with low priority tasks.

\section{Conclusion}\label{sec:conclusion}
In this work we consider the problem of independent learning in mixed-motive games. To mitigate the inefficiencies from independent learning we propose a market mechanism, which we formally define as Stochastic Market Game (SMG), in order to allow agents to interact by means of the market (meta) game. Furthermore, two classes of markets are introduced, categorized by the way trades between agents are specified as either conditional or unconditional. We analyze markets in the Prisoner's Dilemma, where market changes the game dynamics to let independent Q-learning consistently achieve optimal, i.e. cooperative results. Furthermore, two variations of a more complex environment are used, where agents face the challenge of partially conflicting goals under the constraint of shared resources and we find that the proposed markets significantly increase overall and agent specific returns.



\bibliographystyle{named}
\bibliography{main}

\end{document}